\documentclass[article,onecolumn]{revtex4}
\usepackage{graphicx}
\usepackage{color}
\definecolor{darkred}{rgb}{0.75,0.0,0.0}
\newcommand\beq{\begin{equation}}
\newcommand\eeq{\end{equation}}
\newcommand\bear{\begin{eqnarray}}
\newcommand\eear{\end{eqnarray}}
\begin{document}

\title {Effect of Electric Field on One-Dimensional Insulators: 
A DMRG study} 
\author{Sudipta Dutta, S. Lakshmi and Swapan K Pati}

\address{Theoretical Sciences Unit and DST unit on Nanoscience
\\Jawaharlal Nehru Centre For Advanced Scientific Research
\\Jakkur Campus, Bangalore 560064, India.}

\date{\today}
\widetext

\begin{abstract}
\parbox{6in}

{We perform density matrix renormalization group (DMRG) calculations 
extensively on one dimensional Mott and Peierls chains with explicit 
inclusion of the static bias to study the insulator-metal transition
in those systems. We find that the electric field induces a number of 
insulator-metal transitions for finite size systems and at the 
thermodynamic limit, the insulating system breaks down into a 
completely conducting state at a critical value of bias which depends 
strongly on the insulating parameters. Our results indicate that
the breakdown, in both the Peierls and Mott insulators, at thermodynamic
limit, does not follow the Landau-Zener mechanism. Calculations on 
various size systems indicate that an increase in the system size 
decreases the threshold bias as well as the charge 
gap at that bias, making the insulator-metal transition sharper in 
both cases.

\bigskip

}
\end{abstract}

\maketitle

\narrowtext

    Strongly correlated low-dimensional electronic systems have 
attracted much interest because of their unique low-energy 
characteristics\cite{Geller,NI,Others,Wiegmann,Ogota}. These
systems are almost always insulators due to various interactions in 
reduced dimensions and are commonly described by Hubbard, Peierls or
related Hamiltonians\cite{Hubbard,Peierls}. 
While the ground state of a half-filled Hubbard system is a Mott insulator,
the electron-lattice interactions lead to Peierls instability in low-dimensional
systems. 

The effect of electric field on such systems has 
generated much interest in recent years due to the practical interest in
tuning their dielectric and piezoelectric properties, more so for the
Mott insulators. Many electronic conduction processes seem to suggest
electric field induced phenomena, such as negative differential resistance
in molecular electronics\cite{Tour,Lakshmi1}. Recent experiments on
low-dimensional Mott insulators, namely, Sr$_2$CuO$_3$, SrCuO$_2$\cite{Taguchi}
and La$_{2-x}$Sr$_x$NiO$_4$\cite{Yamanouchi} suggest a dielectric breakdown 
in presence of an external electric field.
While the Peierls case is relatively easier, a tractable computational 
method which takes into account the static electric field and its response 
on the excited correlated electronic states is still lacking.
One of the earliest theoretical approaches considering
electric field was the Bethe-{\it ansatz} 
method\cite{Fukui,Deguchi}, where, asymmetric tunneling terms 
towards the left and right were considered inorder to describe the 
non-equilibrium situation and hence capture the flavor of an electric 
field. However, the imaginary gauge term giving rise to a 
non-hermitian Hamiltonian makes it very difficult to relate this to a system 
in presence of a real electric field. In order to account for this, 
another approach to this problem has been to employ the time-dependent 
Schr\"{o}dinger equation to study the time evolution of the many-body wave 
functions and levels, where the electric field is applied {\it via}
a time-dependent Aharonov-Bohm flux\cite{ABflux}. However, in this case the 
calculations are restricted to only ring structures (to avoid
ambiguities arising from the electrodes) and for system sizes $\le$ 10 sites.
Recently, however, there have been a number of studies using the real time
evolution of the ground state in the presence of a source drain 
potential\cite{Dagotto}. However, a microscopic understanding of the 
insulator-metal transition and the quantification of the critical static 
electric field, required to induce it in real extended systems is still lacking.

For a clear understanding of the effect of electric field on finite size
systems as well as in the thermodynamic limit,
in this letter, we use the Density Matrix Renormalization Group 
(DMRG)\cite{White,Schollwock} method which is known to be highly 
accurate for low-dimensional interacting systems. For the first time, we have 
included static electric field in the DMRG algorithm and have obtained 
ground and low-energy eigen states of a one-dimensional system 
with various sizes. The static electric field is included as a ramp 
potential and we find that the electric field can induce a number of
insulator-metal transitions for finite size systems, with the transition 
depending on the Hamiltonian parameters. We analyze our results at   
thermodynamic limit in the light of Landau-Zener mechanism\cite{LZ1}, 
for both the Peierls and Mott insulating systems. To get the thermodynamic 
behavior we consider the total bias instead of field.

We consider a one dimensional chain representing a conjugated polymer or
any other one-dimensional system, described by the Peierls-Hubbard 
Hamiltonian,
\begin{eqnarray}
H &=& \sum\limits_{i}(t+(-1)^{i+1}\delta )(a^\dagger{_i}a_{i+1}+h.c) \nonumber \\
&+& U \sum\limits_{i}n_{i\uparrow}n_{i\downarrow}
\end{eqnarray}

\noindent where t is the hopping term, U is the Hubbard term and $\delta$ 
is the bond alternation parameter. We set $t=1$ and for the Mott-insulator, 
we consider $\delta=0$ and nonzero $U$, while for the Peierls-insulator, 
$U$ is set to zero with nonzero $\delta$. The external electric field applied 
on the system has the form of a ramp potential, distributed over all the 
sites in such a way that the potential $V_{i}$ at site $i$ becomes 
$-\frac{V}{2}+i\frac{V}{N+1}$, where V is the applied voltage and $N$ 
is the total number of sites in the 1D chain. This form of the potential 
ensures that the bias varies between $-V/2$ to $V/2$ across the molecule. 
The potential adds an extra term $\sum\limits_{i}V_{i}a^\dagger{_i}a{_i}$ 
to the above Hamiltonian.

Our DMRG results compare fairly well with the $V=0$ Bethe-{\it ansatz} 
ground state energy and charge gap, for all values of $U$, with a 
density-matrix cut-off $m = 140$. For nonzero bias, the DMRG results compare up to 
numerical accuracies with the exact diagonalization results of finite sizes 
up to $N=16$. For $U=0$, the problem can be exactly solved and we obtain the 
ground state and excitation spectrum in presence of bias using 
tight-binding one-electron formalism. In the Mott (Peierls) case, 
$U(\delta)$ has been varied from 0 to 5 (0 to 1) with bias from 0 to 6 volts in 
steps of $0.1$ volt.

\begin{figure}
\centering
\includegraphics[scale=0.3, angle=0] {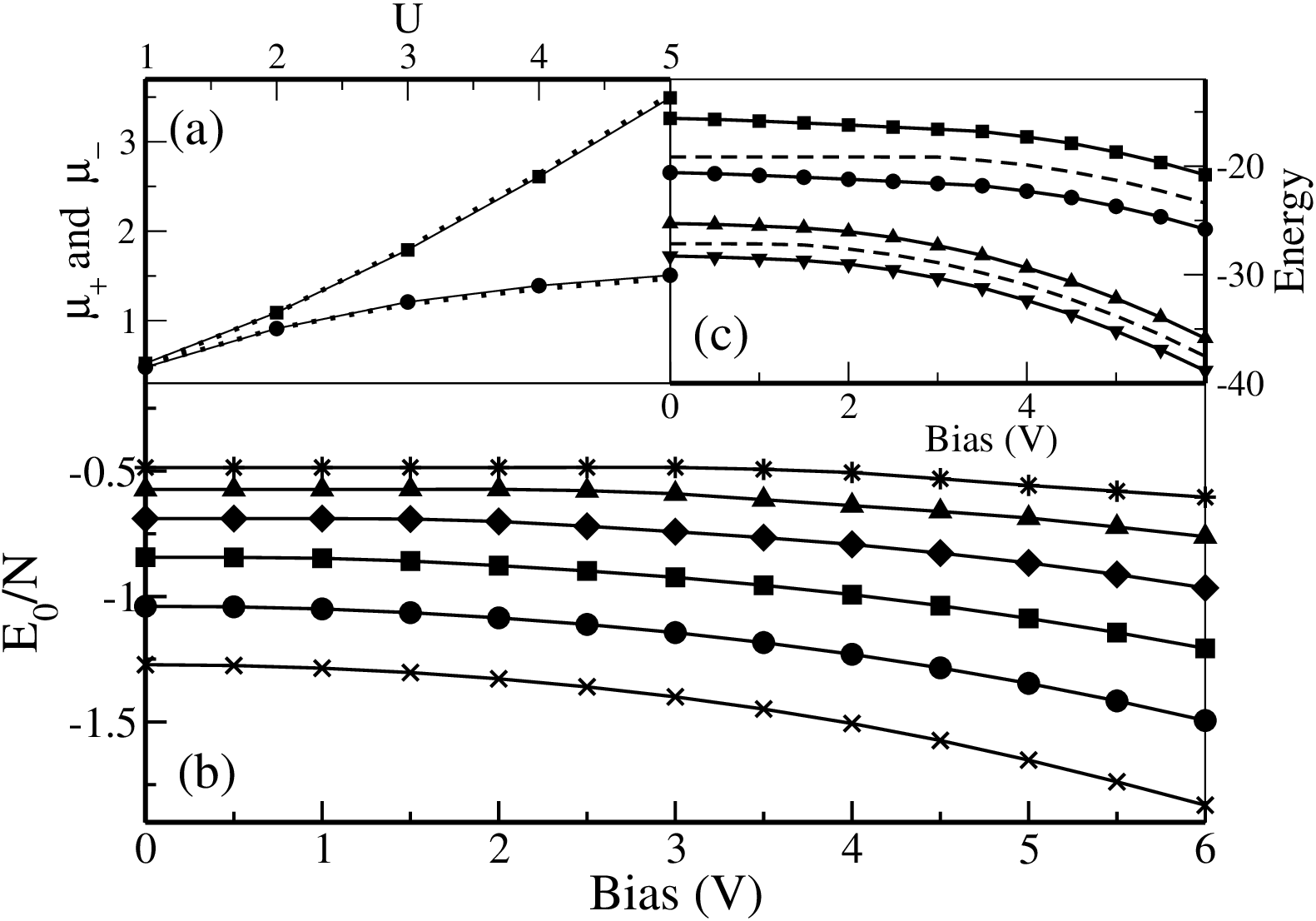}
\caption{(a)$N\rightarrow\infty$ values of $\mu_{+}$(square) and 
$\mu_{-}$(circle) at zero bias as a function of U. The dotted lines 
correspond to exact Bethe-{\it ansatz} results. (b)Converged ground state 
energy per site $\frac{E_{0}}{N}$ for half-filled system (N electrons) vs 
applied bias $(V)$ for $U=$ 0 (X), 1 (circle), 2 (square), 3 (diamond), 
4 (triangle) and 5(star). (c) shows the energy of the systems with N+1 
(up triangle (square)), and N-1 (down triangle(circle)) electrons for 
$U=3 (5)$ for a chain of $40$ sites as a function of bias. The dashed lines 
represent the corresponding ground state energy (E(N)).}
\end{figure}

For many-body models, the charge excitation gap is defined as the difference
between the energy required to add ($\mu_+$) and remove ($\mu_-$) electrons from
the ground state\cite{insulator},
\begin{eqnarray}
\Delta_{charge}=\mu_{+}-\mu_{-}
\end{eqnarray}

\noindent where $\mu_{+}=E(N+1)-E(N)$ and $\mu_{-}=E(N)-E(N-1)$.
$E(N)$, $E(N+1)$ and $E(N-1)$ are the energies of the half-filled system and
the systems with one extra and one less electron respectively.

To understand the energy cost due to the addition or removal of
electrons in the presence of bias, we have calculated $\mu_{+}$ and
$\mu_{-}$ numerically using DMRG for a range of bias. In Fig.1(a),
we have plotted $\mu_{+}$ and $\mu_{-}$ as a function of $U$ at
zero bias. The finite size DMRG results ($\mu_{+}$
and $\mu_{-}$) are extrapolated to the thermodynamic limit
($N\rightarrow\infty$) for every $U$ value. For a clear demonstration
of the accuracy of our results, we have also shown the Bethe-{\it ansatz}
$\mu_{+}$ and $\mu_{-}$ values derived for every $U$ in the same plot.
As can be seen, our numerical extrapolated values compare fairly
well with the exact thermodynamic Bethe-{\it ansatz} results.
Both $\mu_+$ and $\mu_-$ increase
with increase of $U$. However, the former increases with larger slope
than the latter, resulting in increase of the charge gap with
increasing Hubbard repulsion. This is because, a higher value of Hubbard
repulsion localizes the electrons more and leads to higher charge gap.

\begin{figure}
\centering
\includegraphics[scale=0.3, angle=0] {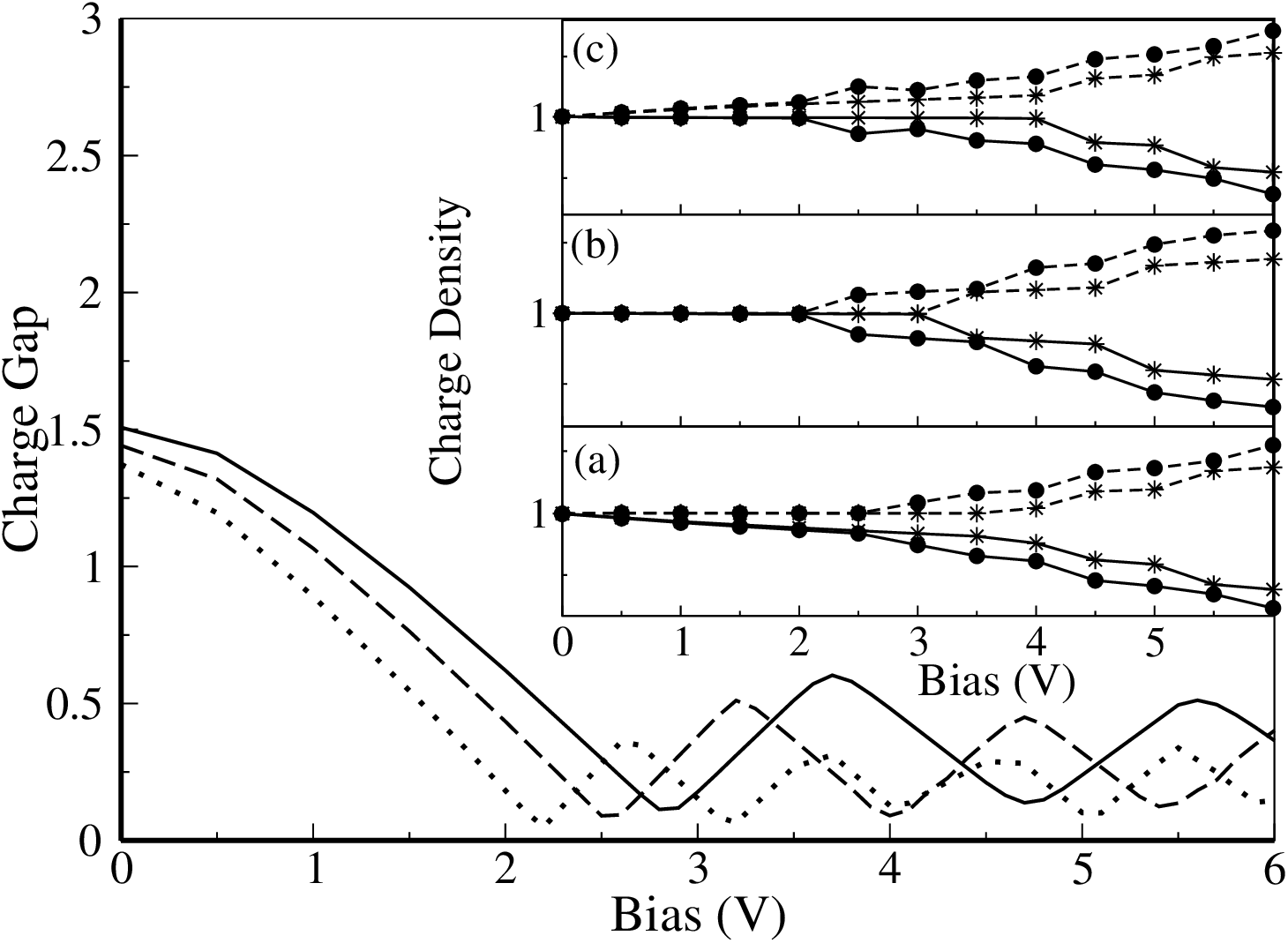}
\caption{Charge gap vs bias for $N=20$(solid line), $26$(dashed line)
and $40$(dotted line) for $U=4$. Inset shows the charge densities on
first (solid lines) and last site (dashed lines) for the N-1(a),
N(b) and N+1(c) electron systems as a function of bias for $U=4$(circle)
and $5$(star) with $N=30$.}
\end{figure}

The ground state of the Hamiltonian with nonzero $U$ is a spin-density wave
insulator with one electron at every site\cite{Sumit}. In Fig.1(b) we have 
plotted the converged (extrapolated value at $N\rightarrow\infty$ limit) 
ground state energy of the half-filled system as a 
function of applied bias for different $U$ values. As can be seen, for a 
fixed $U$ value, the ground state energy remains almost constant upto a 
certain bias, and beyond it, the system starts stabilizing rapidly with 
applied bias\cite{Ivo}. It can thus be interpreted from Fig.1(b) that the 
applied electric field stabilizes the system after a threshold value, 
required to overcome the effect of the electron repulsion, $U$. 
Thereafter, the system gains kinetic energy as the electrons start hopping 
in the direction of bias. For $U=0$, the stabilization starts 
as soon as the bias is turned on. 

To understand the response of the excited states of the system with electric
field, for a given $U$, we have computed $E(N+1)$ and $E(N-1)$.
Both the energies for a particular system size $(N=40)$ are plotted in the 
Fig.1(c) as a function of bias for two representative $U$ values, 
together with $E(N)$. As like the
ground state, excitation energies (both $E(N+1)$ and $E(N-1)$) decrease 
with increase in bias, however, their difference always remains equal to $U$ 
for all values of bias, as expected. For a given value of 
$U$, although the slope of $E(N+1)$ and $E(N-1)$ are same, the slope
of $E(N)$ is different leading to the interesting phenomenon of 
insulator-metal transition, as described later. Fig.1(c) also shows that the 
stabilization of the system with one extra/less electron needs higher 
bias for higher value of $U$. 

\begin{figure}
\centering
\includegraphics[scale=0.3, angle=0] {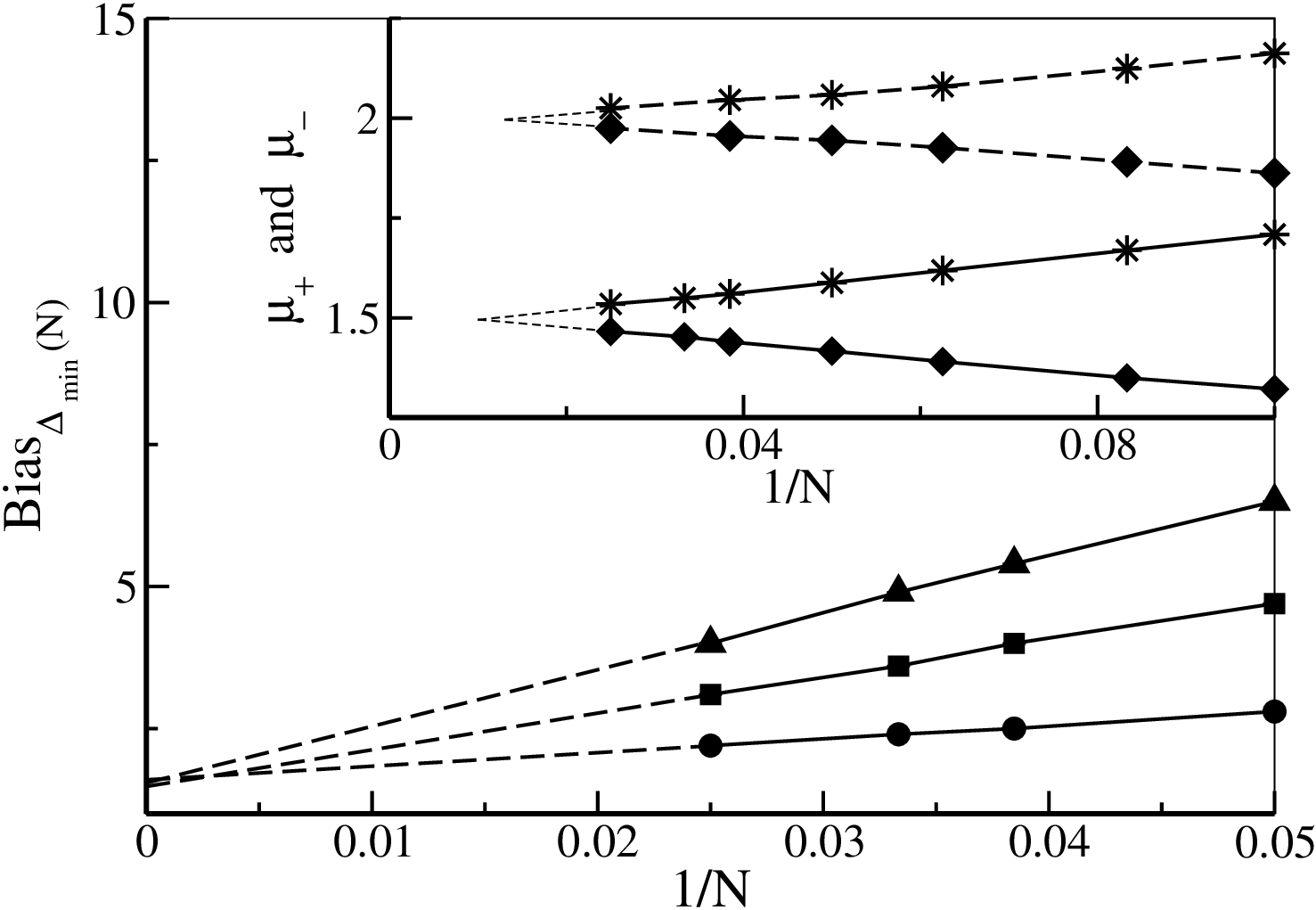}
\caption{Bias at first (circle), second (square) and third (triangle) minimum
as a function of $1/N$ for $U=4$. Dashed lines show the extrapolation. Inset shows the
$\mu_+$(star) and $\mu_-$(diamond) at first minimum as a function of $1/N$ for
$U=3$(solid lines) and $4$(dashed lines). Dotted lines are extrapolation. Error bars
are of the same size as the symbols.}
\end{figure}

To characterise the bias at which this insulator-metal transition 
occurs accurately, we plot in Fig.2 the charge gap as a function 
of bias for different finite size 
systems for a representative value of $U$. It can be clearly seen that 
the charge gap shows an oscillation with bias, going through a number of
minima and maxima.
To understand the underlying physics, we have calculated the onsite charge 
densities as a function of bias for 
various system sizes with several $U$ values. We present this in the inset 
of Fig.2 for the half-filled state and the states with one extra
and one less electron than half-filling for two representative values of
$U$ for a finite chain with $N=30$. For clarity, we plot charge 
densities at only the first and the last sites. At zero bias, the ground 
state charge density at every site of the system is the same and it remains 
almost the same with increase in bias upto the bias corresponding to the first 
minimum of the charge gap ($\Delta E$). However, after that, they show a 
large shift in 
the direction of bias, giving rise to charge inhomogeneties. A higher value 
of $U$ requires a higher bias to shift the charge density as can be seen from 
the inset of Fig.2 and thus the bias corresponding to the first $\Delta E$ minimum 
increases with increasing $U$. Interestingly, the repetetive period of charge 
gap going through minima with increase in bias is due to the role of charge
stiffness. The external bias tends to shift the charge densities towards
one electrode with the nullification of $U$ at the first $\Delta E$ minimum.
However, beyond this, an increase in bias results in further hopping of 
charges leading
to double occupancy of more sites, with electron repulsion overwhelming the
kinetic stablization, thereby increasing the energy gap. Further increase of bias
nullifies this effective repulsion, resulting in the next charge gap 
minimum. Hence, such a variation in charge gap resulting in near-metallic 
(charge gap is not zero) behavior in various bias
regions is due to the interplay of the Hubbard repulsion, finite system size and
the spatial gradient of the external bias. Interestingly, as can be seen from Fig.2, 
with increase in the size, the magnitude of bias corresponding to the
first $\Delta E$ minimum reduces and the periodicity of occurence of
succesive minima thereafter also narrows down. The field $(V/N)$ required to reach
the first $\Delta E$ minimum decreases more sharply with increase in system size.

To explore the size dependence of this insulator-metal transition, in Fig.3 
we plot the bias values corresponding to the first, second and 
third $\Delta E$ minimum separately as a function of $1/N$ and extrapolate the plots
to $N\rightarrow\infty$ limit. All the minima converge to the same point at 
the thermodynamic limit, clearly indicating that at $N\rightarrow\infty$ 
limit, the system goes to conducting state at some bias which we term as 
critical bias, $V_c$. At this $V_c$, the ground state of 
the system changes its slope and starts stabilizing rapidly as can be seen 
from Fig.1(a). In the inset of Fig.3, we have shown $\mu_+$
and $\mu_-$ at first minimum as a function of inverse system size for $U=3$ and $4$.
With increase in system size, $\mu_+$ and $\mu_-$ come close to each
other and thus the gap at first minimum becomes lower with larger deconfinement.
From our finite size data analysis, with extrapolation to $N\rightarrow\infty$
limit, we find that with increase in Hubbard repulsion term, $U$, the 
difference between $\mu_+$ and $\mu_-$ at first minimum for a particular finite 
chain decreases, as can be seen from the inset. Thus with increase in $U$, 
the length of the chain at which the charge gap vanishes, 
decreases. Hence, as is evident from the inset, for a given value of $U$, our 
method enables us to predict the length of the chain at which the system will 
become completely conducting.

\begin{figure}
\centering
\includegraphics[scale=0.29, angle=0] {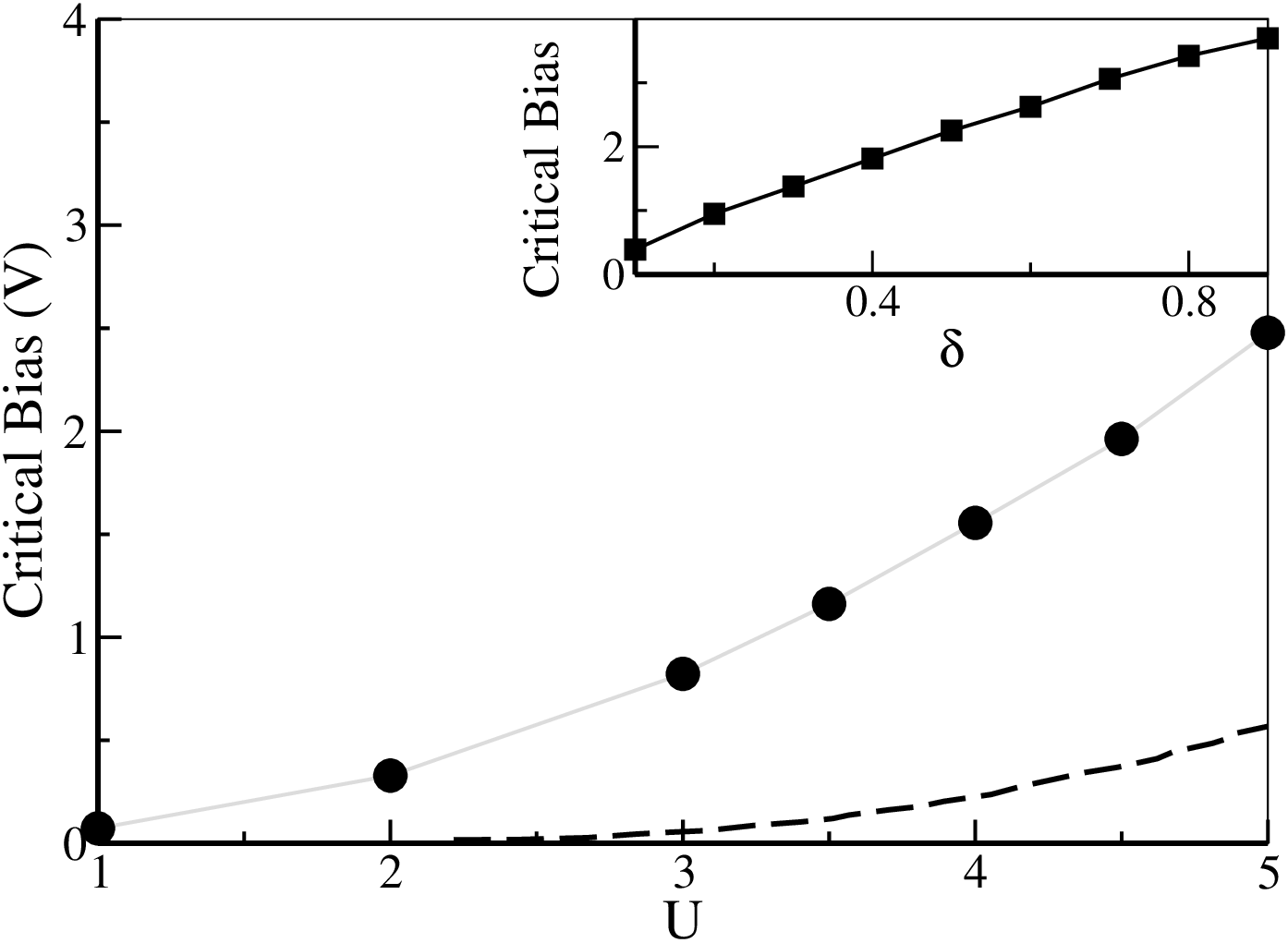}
\caption{$V_c$ as a function of $U$ from our results (circle), and
Bethe-{\it ansatz} prediction of Landau-Zener tunneling (dashed line).
Inset shows $V_c$ as a function of $\delta$.}
\end{figure}

It is clear from the above discussion that, the bias required for the 
closing of Mott gap depends sensitively on the value of $U$. To understand 
how $V_c$ depends on $U$, we plot in Fig.4, the $V_c$ as a function of $U$ obtained
from our calculations, along with that calculated using the Landau-Zener 
formula\cite{Oka} on the initial charge gaps obtained using the Bethe-{\it ansatz}.
Fig.4 shows non-linear dependence 
of $V_c$ on $U$. We have also analyzed the insulator-metal
transition in case of Peierls insulators. In this case, the difference 
between the highest occupied and lowest unoccupied single electron states 
(opposite parity) is the charge gap. Here too, the critical bias has a 
sensitive dependence on the bond-alternation parameter, $\delta$, as has 
been described previously by us\cite{Lakshmi2,Lakshmi3}. Inset of Fig.4 displays the 
variation of converged $V_c$ with $\delta$. 
To now compare and contrast the critical bias obtained for the Peierls and Mott 
insulators in light of the Landau-Zener formula, we analyze the initial charge gap
and the $V_c$ dependence on the model parameters, $\delta$ and $U$. For Peierls
insulators, the $V=0$ charge gap is linear in $\delta$, and the critical bias $V_c$, 
calculated at the thermodynamic limit (inset in Fig. 4) has a near linear dependence 
on $\delta$. This indicates that the Landau-Zener formula which has the form
$V_c(\delta) \propto [\Delta_{charge}(\delta)]^2$, is not obeyed although 
previous studies on finite size systems indicated that tunneling occurs between 
the Landau quasi-degenerate valence and conduction levels\cite{TOka,Lakshmi2,Lakshmi3},
resulting in an exchange of their symmetries. This is because, at the large $N$ 
limit, the gap at the critical bias goes to zero and the system goes
to a completely conducting state. In case of the Mott insulator, the initial 
charge gap has a nonlinear dependence on $U$, and as seen in Fig. 4, the 
critical bias also shows a nonlinear dependence on
$U$. A fitting gives $V_c(U) \propto [\Delta_{charge}(U)]^n$, with $n \sim 1$,
which, unlike the results from previous studies\cite{Oka} clearly indicates that,
the tunneling between many body levels does not follow Landau-Zener mechanism
in thermodynamic limit.
Note that, in our case $V_c$ corresponds to the total bias applied to the chain with
$N \rightarrow \infty$, while the time dependent studies\cite{Oka,TOka} calculate the
critical field $(V_c/N)$ by converging it upto a certain finite chain length.
We do not consider the effect of polarization on the applied electric field 
in our calculations. Because the inclusion of polarization does not change the 
nature of the insulator-metal transition as we have observed earlier\cite{Lakshmi2}.
It can only change the quantitaive estimation of the bias at charge gap minima to 
some extent. Moreover, in presence of Hubbard repulsion term $U$ the 
ramp nature of the electric field is retained even with inclusion of 
polarization effects in our calculations\cite{Lakshmi4}.

Qualitatively, the slope change of the ground state
can be visualized even for a $2$-sites Mott-insulating system. For $V=0$, 
the ground state of the half-filled system comprising of four basis states, 
has large contibution from the singly occupied sites, with second order ($t^2/U$) 
contribution from the double occupancy sites. However, the external bias nullifies 
the Hubbard repulsion, leading to a mixing of the double and single occupancy sites. 
The ground state charge density shifts in the direction of the applied electric 
field, at $V \sim 3U$. On the other hand, the state with one extra electron comprising 
of two basis states, stabilizes with a slope which does not 
depend on $U$. This is because the external bias results in the hopping of electron
from one site to another, leading to a preference of one of the basis states
as compared to the other. But since both the basis states are energetically 
degenerate, this hopping occurs without any Hubbard energy cost, rather the
energy reduces due to kinetic stabilization. This results in a slope 
that is independent of the Hubbard repulsion. Additionally, the Hubbard
repulsion also does not play any role on the $2$-sites one electron case.
The $2$-sites problem thus captures the ground state behavior, and not the
transition pheonomena, since the interesting features described earlier 
($\Delta E$ oscillation) leading to breakdown behavior occur primarily due
to the response of the excited states to an applied static electric field.

In conclusion, we have shown that, in thermodynamic limit, the insulator-metal 
transition in presence of an external static electric field does not follow 
Landau-Zener mechanism in both Peierls and Mott insulators. Our DMRG calculation 
is the first of its kind, which allows us to study various system sizes with 
better understanding of the general insulator-metal transition, with quantitative 
predictability.

SD acknowledges the CSIR for the research fellowship and SKP acknowledges 
the research support from DST and CSIR, Govt. of India.

\end{document}